\documentclass[aps,nofootinbib,pra,superscriptaddress,notitlepage]{revtex4-1}

\usepackage[T1]{fontenc}
\usepackage[utf8]{inputenc}
\usepackage{amsmath}
\usepackage{amsthm}
\usepackage{amsfonts,dsfont}
\usepackage{mathrsfs}
\usepackage{graphicx}
\usepackage{bbm}
\usepackage{hyperref}
\hypersetup{linktocpage,colorlinks,citecolor={blue},pdfdisplaydoctitle=true,pdfpagemode=UseOutlines,bookmarksnumbered=true}
\usepackage{color}
\usepackage{comment}

\newcommand\ket[1]{\left|#1\right>}
\newcommand\bra[1]{\left<#1\right|}

\newcommand\Ev{\mathbb{E}}

\newcommand\Tr{\mathrm{Tr}}
\newcommand\ro{{\hat\rho}}
\newcommand\Ho{{\hat H}}
\newcommand\Vo{{\hat V}}
\newcommand\xo{{\hat{\mathbf{x}} }}

\newcommand\kb{{\mathbf{k}}}
\newcommand\xb{{\mathbf{x}}}
\newcommand\Dcal{\mathcal{D}}

\newcommand\Ao{{\hat A}}
\newcommand\Bo{{\hat B}}
 
\newcommand{\upd}[0]{\mathrm{d}}
\newcommand{\ie}[0]{\textit{i.e.} }
\newcommand{\eg}[0]{\textit{e.g.} }

\newcommand{\V}{\mathds{V}}

\begin{document}

\title{On GKLS dynamics for local operations and classical communication}

\author{Lajos Di\'osi}
\email{diosi.lajos@wigner.mta.hu}
\affiliation{Wigner Research Centre for Physics, H-1525 Budapest 114. P.O.Box 49, Hungary}

\author{Antoine Tilloy}
\email{antoine.tilloy@mpq.mpg.de}
\affiliation{Max-Planck-Institut f\"ur Quantenoptik, Hans-Kopfermann-Stra{\ss}e 1, 85748 Garching, Germany}

\date{\today}

\begin{abstract}
We define a time continuous version of the concept of "local operations and classical communication" (LOCC), ubiquitous in quantum information theory. It allows us to construct GKLS master equations for particle systems that have (1) an arbitrary pair potential, and (2) local decoherence terms, but that do not entangle the constituents. The local decoherence terms take a particularly simple form if a principle of least decoherence is applied.
\end{abstract}

\maketitle

\section{Introduction}
Local operations and classical communication (LOCC) are an important subclass of operations on multipartite quantum systems, introduced first in the context of quantum information theory \cite{nielsen2010,diosi2011}. In the bipartite case, two subsystems are possessed by Alice and Bob respectively. They can carry out local operations as well as classically communicate with each other but are not allowed to exchange quantum information (\eg by sending each other photons in definite entangled states). The interest of this class of operations is that it can correlate the local subsystems but it cannot entangle them. For this reason, LOCC are instrumental in basic quantum information protocols, allowing entanglement to be seen as a resource one can gather (entanglement distillation) to subsequently enable tasks (such as quantum teleportation). Despite what their simple definition would suggest, quantum dynamical maps obtained from LOCC have a rich and intricate structure \cite{chitambar2014}. The time continuous setting, which we shall discuss below, has been less studied but already points to many interesting problems.

The motivation for the analysis of continuous time LOCC dynamics is unexpected and arose recently in a context a priori far removed from quantum information theory. Recent works \cite{kafri2014,kafri2015,tilloy2016,altamirano2016,tilloy2017,tilloy2017grw,bose2017} have indeed used continuous time LOCC to construct and constrain semi-classical theories of Newtonian gravity. The idea is that if the gravitational force is classical at a fundamental level, then perhaps it cannot be used to entangle particles. A sufficient condition\footnote{This is only a sufficient condition because there could exist non-entangling dynamics that cannot be simply derived from continuous time LOCC (see \eg the slightly larger class of semi-classical dynamics discussed in \cite{tilloy2017}). Further, the very definition of classical or non-quantum is sufficiently flexible that one may wish to extend it beyond non-entangling dynamics \cite{hall2017}.} is then that gravity can be implemented with LOCC. This provides a way to construct explicit theories of semi-classical gravity that do not require a modification of the statistical toolbox of quantum mechanics contrary to non-linear mean field approaches.

Let us informally define continuous time LOCC dynamics in a simple bipartite setting. In the discrete setting, Alice performs a local measurement on her system and broadcasts the measurement outcome to Bob. Knowing this outcome, Bob then performs a local unitary. These steps can be repeated arbitrarily many times while Alice and Bob can exchange their role. There is a natural continuous time limit to such an iteration of LOCC and one can obtain it using the theory of continuous quantum measurement. Alice continuously measures (or \emph{monitors}) a local observable $\hat{A}$ and broadcasts the time dependent measurement signal $a(t)$ to Bob. Bob then applies in real time a local feedback Hamiltonian $\Vo = a(t)\times \Bo$, proportional to the received signal. This is an instance of Markovian measurement based feedback in which the measured operators are restricted to a single subsystem of a bipartite decomposition. The corresponding Gorini-Kossakowski-Lindblad-Sudarshan (GKLS) master equation \cite{gorini1976,lindblad1976}, even without this locality constraint, was derived a long time ago \cite{caves1987,diosi1990,wiseman1993,wiseman1994,diosi1994,wiseman2009}. Importantly, the feedback master equation contains non-local Hamiltonian terms in addition to local decoherence and possibly non-local dissipation.

This brings a number of natural questions. \emph{What kind of non-local potential can one construct this way? What is the price to pay in local decoherence or equivalently in local classical noise? Is there a natural way to define the minimum amount of noise required to implement a given non-local quantum potential with LOCC dynamics?} Our objective is to answer those questions in the slightly restrictive context of bipartite systems and for translation invariant potentials.

\section{Continuous local monitoring and non local feedback}

\subsection{One-way case}
We consider a bipartite quantum system of state $\hat{\rho}$ with each part under the control of Alice and Bob respectively. In the absence of feedback, we assume that the total Hamiltonian is a sum of local terms $\Ho=\Ho_A+\Ho_B$. We further assume that Alice is monitoring a set of observables $\{\Ao_\nu\}$. The statistics of the corresponding measurement signals $a_\nu(t)$ as well as the resulting backaction on the state can be written down using continuous measurement theory (see \eg \cite{jacobs2006,wiseman2009}). One gets:
\begin{equation}
a_\nu(t)=\Tr(\Ao_\nu\ro_t)+w^A_\nu(t),
\end{equation}
where the $w^A_\nu(t)$ are correlated white noises of zero average characterized by their two point function
\begin{equation}
\Ev\left[w^A_\nu(t) w^A_\mu(s)\right]=(\gamma^A)^{-1}_{\nu\mu}\delta(t-s).
\end{equation}
The real non-negative matrix $\gamma^A$ encodes the precision of Alice's monitoring device. Under this continuous measurement, the state $\ro^w_t$ evolves according to the stochastic master equation (SME):
\begin{equation}
    \frac{\upd\ro^w}{\upd t}=-i[\Ho_A+\Ho_B,\ro^w]
-\frac{1}{2}\sum_{\nu\mu}\left(\frac{\gamma^A_{\nu\mu}}{4}\left[\Ao_\nu,[\Ao_\mu,\ro^w]\right] - \gamma^A_{\nu\mu} \left\{\Ao_\nu - \langle \Ao_\nu\rangle_t, \ro^w_t\right\} w^A_\mu(t) \right)
\end{equation}
where $\langle \Ao_\nu\rangle_t = \Tr [\Ao_\nu\ro^w_t]$ and the multiplicative white noise is understood in the It\^o convention. If nothing is done with the measurement signals and if one averages over them, one gets the simple GKLS master equation for $\ro=\mathds{E}[\ro^w]$:
\begin{equation}
    \frac{\upd\ro}{\upd t}=-i[\Ho_A+\Ho_B,\ro]
-\frac{1}{2}\sum_{\nu\mu}\frac{\gamma^A_{\nu\mu}}{4}\left[\Ao_\nu,[\Ao_\mu,\ro]\right]. 
\end{equation}
However, if Bob introduces a local potential $\Vo_B(t)$ proportional to the instantaneous signals he receives, \ie
\begin{equation}
\Vo_B(t)=\sum_\nu a_\nu(t)\Bo_\nu, 
\end{equation}
where $\{\Bo_\nu\}$ are local observables of Bob,
one can show that the noise-averaged state of the composite system satisfies the
GKLS equation:
\begin{equation}\label{eq:masterasym}
\frac{\upd\ro}{\upd t}=-i[\Ho_A+\Ho_B,\ro]
-\frac{1}{2}\sum_{\nu\mu}\left(i\left[\Bo_\nu,\{\Ao_\nu,\ro\}\right]
+\frac{\gamma^A_{\nu\mu}}{4}\left[\Ao_\nu,[\Ao_\mu,\ro]\right]
+(\gamma^A)^{-1}_{\nu\mu}\left[\Bo_\nu,[\Bo_\mu,\ro]\right]\right).
\end{equation}
We shall not discuss the derivation of this master equation which contains minor subtleties and which can be found in the literature (see \eg \cite{wiseman1994,wiseman2009,tilloy2016}). Rather, we take \eqref{eq:masterasym} as the starting point of our exploration of continuous time LOCC dynamics. We note already an interesting feature of the ME \eqref{eq:masterasym}: the non-Hamiltonian part is non-local due to the terms $-\frac{i}{2}[\Bo_\nu,\{\Ao_\nu,\ro\}]$ although the equation is non-entangling by construction, as it is obtained from LOCC dynamics. 

\subsection{Symmetric case} 
We extend the previous setting by adding the possibility for Bob to measure and Alice to apply a potential proportional to the corresponding instantaneous signal. The general case can be obtained in the same way as before. We focus on a restricted symmetric case which is particularly interesting. Let us assume that Bob is monitoring the same set of observables $\{\Bo_\nu\}$ that he is using for feedback and that Alice is using the same set of observables $\{\Ao_\nu\}$ for feedback that she is using for monitoring. The corresponding GKLS master equation --after averaging over the measurement outcomes-- reads:
\begin{equation}\label{eq:mastersym}
\frac{\upd\ro}{\upd t}=-i[\Ho_A+\Ho_B+\sum_\nu\Ao_\nu\Bo_\nu,\ro]
-\sum_{\nu\mu}\left(\frac{\gamma^A_{\nu\mu}}{8}+\frac{(\gamma^B)^{-1}_{\nu\mu}}{2}\right)\left[\Ao_\nu,[\Ao_\mu,\ro]\right]
-\sum_{\nu\mu}\left(\frac{\gamma^B_{\nu\mu}}{8}+\frac{(\gamma^A)^{-1}_{\nu\mu}}{2}\right)\left[\Bo_\nu,[\Bo_\mu,\ro]\right],
\end{equation}
where the only freedom lies in the two non-negative real matrices $\gamma^A$ and $\gamma^B$ setting the precision of Alice's and Bob's monitoring. The master equation \eqref{eq:mastersym} possesses two remarkable features. As before, it is obtained from LOCC and thus it does not entangle the two parties. Second, it contains a non-local Hamiltonian term:
\begin{equation}\label{eq:H_AB}
\Ho_{AB}=\sum_\nu\Ao_\nu\Bo_\nu.
\end{equation}
The coexistence of these two properties is made possible by the presence of local decoherence on both sides. Because every interaction Hamiltonian can be decomposed into a sum of tensor products of local terms, we answer the first question mentioned in the introduction: provided one accepts to pay a certain price in local decoherence, one can implement any non-local Hamiltonian on a bipartite system\footnote{The only subtlety, as we shall later see, is that implementing a given Hamiltonian may require divergent local decoherence terms making the corresponding master equation trivial.}. We now discuss more precisely the cost in local decoherence to implement a given pair potential between two particles.

\section{Implementing a pair potential}

\subsection{Setup}\label{sec:setup}
We assume that Alice and Bob each have a particle and that the two particles interact via a translation invariant pair potential $V(\xo_A-\xo_B)$ depending only on the particle coordinates. Naturally, even for initially disentangled states, the pair-potential will entangle the bipartite system. If, however, there is sufficiently strong local decoherence, then entanglement vanishes in finite time and the two particles remain only classically correlated \cite{diosi2003,yu2004}. We assume that the bipartite state $\hat{\rho}$ obeys the GKLS equation:
\begin{equation}\label{eq:localdeco}
\frac{\upd\ro}{\upd t}=-i\left[\Ho_A+\Ho_B+V(\xo_A-\xo_B),\ro\right]+\Dcal_A\ro+\Dcal_B\ro,
\end{equation}
where $\Ho_A$, $\Ho_B$ are local Hamiltonians and $\mathcal{D}_A$, $\mathcal{D}_B$ are local GKLS ``dissipators''. In light of the previous section, the natural question we may want to answer is the following. What structure and ``intensity'' need the local dissipator have to prevent the pair potential from entangling the two particles? A strategy to obtain a candidate is to construct a continuous time LOCC dynamics giving rise to a master equation of the form \eqref{eq:localdeco} and then to minimize the local dissipators. 

We first write the pair potential as a sum of tensor products of local operators by Fourier expanding it. Because the potential is self-adjoint, we get:
\begin{align}
    V(\xo_A-\xo_B) &= \int_{\mathds{R}^3} \upd \kb\; v_\kb \cos\left[\kb\cdot(\xo_A-\xo_B)\right]\\
    &=\int_{\mathds{R}^3} \upd \kb\; v_\kb \left(\cos(\kb\cdot\xo_A)\cos(\kb\cdot\xo_B)+\sin(\kb\cdot\xo_A)\sin(\kb\cdot\xo_B)\right)
\end{align}
where $v_\kb$ is real. Defining now:
\begin{equation}\label{AksBks}
\Ao_{\kb,s}=\sqrt{|v_\kb|}\times\left\{
\begin{array}{c}
\cos(\kb\cdot\xo_A)~~~~s=1\\
\sin(\kb\cdot\xo_A)~~~~s=2
\end{array}
\right.
~~~~~~~~
\Bo_{\kb,s}=\mathrm{sgn}(v_\kb)\sqrt{|v_\kb|}\times\left\{
\begin{array}{c}
\cos(\kb\cdot\xo_B)~~~~s=1\\
\sin(\kb\cdot\xo_B)~~~~s=2
\end{array}
\right. ,
\end{equation}
we can identify the potential with the interaction Hamiltonian \eqref{eq:H_AB} of \eqref{eq:mastersym}:
\begin{equation}\label{V_expansion}
V(\xo_A-\xo_B)=\sum_{s=1,2}\int_{\mathds{R}^3} \upd \kb \; \Ao_{\kb,s}\Bo_{\kb,s}.
\end{equation}
Alice's dissipator reads:
\begin{equation}
\Dcal_A\ro=-\sum_{s,s'=1,2}\iint \upd \kb\upd \kb'\left(\frac{\gamma^A_{\kb,s,\kb',s'}}{8}+\frac{(\gamma^B)^{-1}_{\kb,s,\kb',s'}}{2}\right)\left[\Ao_{\kb,s},[\Ao_{\kb',s'},\ro]\right],
\end{equation}
and Bob's can be written in the same way by exchanging $A$ and $B$.

\subsection{Principle of least decoherence}
To find a way to canonically minimize local decoherence, we make the assumption that the dissipators in the final GKLS equation have the same symmetries as the potential, \ie that they are translation invariant and that they have the $A\leftrightarrow B$ symmetry. This drastically reduces the freedom in the precision matrix which can be written:
\begin{equation}
\gamma^A_{\kb,s,\kb',s'} = \gamma^B_{\kb,s,\kb',s'} = \gamma_{\kb}\,\delta_{s,s'}\,\delta^3(\kb-\kb').
\end{equation}
With this simplification, Alice's dissipator becomes:
\begin{equation}
\Dcal_A\ro=-\sum_{s=1,2}\int \upd \kb\left(\frac{\gamma_\kb}{8}+\frac{1}{2\gamma_\kb}\right)\left[\Ao_{\kb,s},[\Ao_{\kb,s},\ro]\right].
\end{equation}
We now have a clear principle of least decoherence (analogous the the one already discussed in \cite{tilloy2017}) which consists in minimizing all the coefficients independently for each wave number. This yields $\gamma_\kb=2$. In the end the Alice's local decoherence term reads:
\begin{align}
    \mathcal{D}_A \ro&=-\frac{1}{2} \int \upd \kb \,|v_\kb| \left(\left[\cos(\kb\cdot \xo_A),[\cos(\kb\cdot \xo_A),\hat{\rho}]\right]+\left[\sin(\kb\cdot \xo_A),[\sin(\kb\cdot \xo_A),\hat{\rho}]\right]\right)\\
    &= \iint \V(\xb-\xb') \delta^3(\xb-\xo_A)\, \ro\,  \delta^3(\xb'-\xo_A) \,\upd \xb\upd \xb' -\V(0)\,\ro, \label{eq:dissipator}
\end{align}
where $\V$ is the potential with $|v_\kb|$ as Fourier modes (instead of $v_\kb$ for the initial pair potential). If $V$ is a non-negative kernel (as is the case for most non-relativistic Green's functions), then $\V=V$. This is a remarkable result: the minimum of decoherence is reached when the spatial local decoherence functional is given by the pair-potential. This brings a severe difficulty for pair-potentials diverging in $0$ (as is the case for the Newtonian potential) as the resulting minimum decoherence is then infinite. Implementing such a potential in an LOCC way requires a regularization at short distances to keep decoherence finite (see \cite{tilloy2017} for a discussion in the context of gravity). 

\subsection{Minimal disentangling local noise}
The previous results apply in a slightly different context than the one of \ref{sec:setup}. Consider now that he two particles are subjected to local noisy potentials in addition to the entangling pair potential. We assume that their evolution is given by the following stochastic Schr\"odinger equation:
\begin{equation}\label{eq:sse}
\frac{\upd}{\upd t} \ket{\psi}=
-i\left(\Ho_A+\Ho_B+V(\xo_A-\xo_B)+\xi_A(\xo_A,t)+\xi_B(\xo_B,t)\right)\ket{\psi}
\end{equation}
where $\xi_A$ and $\xi_B$ are independent white Gaussian noises of zero mean with the same two point function
\begin{equation}
\Ev\left[\xi_A(x,t)\xi_A(x',t')\right]=\Ev\left[\xi_B(x,t)\xi_B(x',t')\right]=D(x-x')\delta(t-t'),
\end{equation}
and where the multiplicative noise is understood in the Stratonovich convention\footnote{This convention is natural in this context because the noise is physical. Physically, equation \eqref{eq:sse} is obtained in the limit where the time correlation of the noise is much smaller than all other timescales. By virtue of the Wong-Zakaï theorem \cite{wong1965}, this limit corresponds to the stochastic differential equation in Stratonovich form.}. The only constraint on $D$ is that it is a real function corresponding to a non-negative kernel (that is, $D$ has positive Fourier transform). We may ask what amount of local noise is sufficient to destroy the entangling properties of the pair potential. This can be done simply by noting that the GKLS master equation obtained from \eqref{eq:sse}, after averaging, has local dissipators of the same form as those of equation \eqref{eq:dissipator}. Indeed, writing $\ro=\mathds{E}[\ket{\psi}\bra{\psi}]$ one can show\footnote{This can be done by rewriting \eqref{eq:sse} in the It\^o convention, using It\^o's lemma to get a stochastic master equation for $\ket{\psi}\bra{\psi}$, and finally removing the It\^o integral by averaging. Equivalently one may solve \eqref{eq:sse} by formal exponentiation in the Stratonovich representation, average over the noise by Gaussian integration, and differentiate the result.} that $\ro$ obeys the GKLS master equation \eqref{eq:localdeco} with:
\begin{equation}
     \mathcal{D}_A \ro= \iint D(\xb-\xb') \delta^3(\xb-\xo_A)\, \ro\,  \delta^3(\xb'-\xo_A) \,\upd \xb\upd \xb' -D(0)\,\ro,
\end{equation}
and the same for $B$. Identifying this expression with \eqref{eq:dissipator} we naturally fix $D=\V$, \ie the correlation function of the local noises is given by the pair potential (or a version with positive Fourier transform were it not non-negative).

This gives us the noise threshold for which the stochastic Schr\"odinger equation \eqref{eq:sse} is non-entangling (on average). As an aside, this provides a simple demonstration that many-body dynamics are easy to simulate classically provided a sufficient amount of local classical noise is added.

\section{Summary}

We have shown that it is possible to create any potential --not necessarily harmonic-- between two particles using LOCC. The corresponding GKLS master equations possess local decoherence terms that take a particularly simple form provided a principle of least decoherence is applied. In that case the local decoherence functional is simply equal to the pair potential (if the latter has a non-negative Fourier transform). 

\begin{acknowledgments}
L.D. thanks the Hungarian Scientific Research Fund under
Grant No. 124351, and the EU COST Actions MP1209 and CA15220 for support. A.T. thanks the Alexander von Humboldt foundation and the Agence Nationale de la Recherche (ANR) contract ANR-14-CE25-0003-01 for support.
\end{acknowledgments}

\bibliography{main}

\begin{thebibliography}{23}%
\makeatletter
\providecommand \@ifxundefined [1]{%
 \@ifx{#1\undefined}
}%
\providecommand \@ifnum [1]{%
 \ifnum #1\expandafter \@firstoftwo
 \else \expandafter \@secondoftwo
 \fi
}%
\providecommand \@ifx [1]{%
 \ifx #1\expandafter \@firstoftwo
 \else \expandafter \@secondoftwo
 \fi
}%
\providecommand \natexlab [1]{#1}%
\providecommand \enquote  [1]{``#1''}%
\providecommand \bibnamefont  [1]{#1}%
\providecommand \bibfnamefont [1]{#1}%
\providecommand \citenamefont [1]{#1}%
\providecommand \href@noop [0]{\@secondoftwo}%
\providecommand \href [0]{\begingroup \@sanitize@url \@href}%
\providecommand \@href[1]{\@@startlink{#1}\@@href}%
\providecommand \@@href[1]{\endgroup#1\@@endlink}%
\providecommand \@sanitize@url [0]{\catcode `\\12\catcode `\$12\catcode
  `\&12\catcode `\#12\catcode `\^12\catcode `\_12\catcode `\%12\relax}%
\providecommand \@@startlink[1]{}%
\providecommand \@@endlink[0]{}%
\providecommand \url  [0]{\begingroup\@sanitize@url \@url }%
\providecommand \@url [1]{\endgroup\@href {#1}{\urlprefix }}%
\providecommand \urlprefix  [0]{URL }%
\providecommand \Eprint [0]{\href }%
\providecommand \doibase [0]{http://dx.doi.org/}%
\providecommand \selectlanguage [0]{\@gobble}%
\providecommand \bibinfo  [0]{\@secondoftwo}%
\providecommand \bibfield  [0]{\@secondoftwo}%
\providecommand \translation [1]{[#1]}%
\providecommand \BibitemOpen [0]{}%
\providecommand \bibitemStop [0]{}%
\providecommand \bibitemNoStop [0]{.\EOS\space}%
\providecommand \EOS [0]{\spacefactor3000\relax}%
\providecommand \BibitemShut  [1]{\csname bibitem#1\endcsname}%
\let\auto@bib@innerbib\@empty
\bibitem [{\citenamefont {Nielsen}\ and\ \citenamefont
  {Chuang}(2010)}]{nielsen2010}%
  \BibitemOpen
  \bibfield  {author} {\bibinfo {author} {\bibfnamefont {M.~A.}\ \bibnamefont
  {Nielsen}}\ and\ \bibinfo {author} {\bibfnamefont {I.~L.}\ \bibnamefont
  {Chuang}},\ }\href@noop {} {\emph {\bibinfo {title} {Quantum Computation and
  Quantum Information}}}\ (\bibinfo  {publisher} {Cambridge University Press,
  Cambridge UK},\ \bibinfo {year} {2010})\BibitemShut {NoStop}%
\bibitem [{\citenamefont {Di{\'o}si}(2011)}]{diosi2011}%
  \BibitemOpen
  \bibfield  {author} {\bibinfo {author} {\bibfnamefont {L.}~\bibnamefont
  {Di{\'o}si}},\ }\href@noop {} {\emph {\bibinfo {title} {A short course in
  quantum information theory}}},\ Vol.\ \bibinfo {volume} {827}\ (\bibinfo
  {publisher} {Springer, Berlin},\ \bibinfo {year} {2011})\BibitemShut
  {NoStop}%
\bibitem [{\citenamefont {Chitambar}\ \emph {et~al.}(2014)\citenamefont
  {Chitambar}, \citenamefont {Leung}, \citenamefont {Man{\v{c}}inska},
  \citenamefont {Ozols},\ and\ \citenamefont {Winter}}]{chitambar2014}%
  \BibitemOpen
  \bibfield  {author} {\bibinfo {author} {\bibfnamefont {E.}~\bibnamefont
  {Chitambar}}, \bibinfo {author} {\bibfnamefont {D.}~\bibnamefont {Leung}},
  \bibinfo {author} {\bibfnamefont {L.}~\bibnamefont {Man{\v{c}}inska}},
  \bibinfo {author} {\bibfnamefont {M.}~\bibnamefont {Ozols}}, \ and\ \bibinfo
  {author} {\bibfnamefont {A.}~\bibnamefont {Winter}},\ }\href {\doibase
  10.1007/s00220-014-1953-9} {\bibfield  {journal} {\bibinfo  {journal}
  {Commun. Math. Phys.}\ }\textbf {\bibinfo {volume} {328}},\ \bibinfo {pages}
  {303} (\bibinfo {year} {2014})}\BibitemShut {NoStop}%
\bibitem [{\citenamefont {Kafri}\ \emph {et~al.}(2014)\citenamefont {Kafri},
  \citenamefont {Taylor},\ and\ \citenamefont {Milburn}}]{kafri2014}%
  \BibitemOpen
  \bibfield  {author} {\bibinfo {author} {\bibfnamefont {D.}~\bibnamefont
  {Kafri}}, \bibinfo {author} {\bibfnamefont {J.~M.}\ \bibnamefont {Taylor}}, \
  and\ \bibinfo {author} {\bibfnamefont {G.~J.}\ \bibnamefont {Milburn}},\
  }\href {\doibase https://doi.org/10.1088/1367-2630/16/6/065020} {\bibfield
  {journal} {\bibinfo  {journal} {New J. Phys.}\ }\textbf {\bibinfo {volume}
  {16}},\ \bibinfo {pages} {065020} (\bibinfo {year} {2014})}\BibitemShut
  {NoStop}%
\bibitem [{\citenamefont {Kafri}\ \emph {et~al.}(2015)\citenamefont {Kafri},
  \citenamefont {Milburn},\ and\ \citenamefont {Taylor}}]{kafri2015}%
  \BibitemOpen
  \bibfield  {author} {\bibinfo {author} {\bibfnamefont {D.}~\bibnamefont
  {Kafri}}, \bibinfo {author} {\bibfnamefont {G.~J.}\ \bibnamefont {Milburn}},
  \ and\ \bibinfo {author} {\bibfnamefont {J.~M.}\ \bibnamefont {Taylor}},\
  }\href {\doibase https://doi.org/10.1088/1367-2630/17/1/015006} {\bibfield
  {journal} {\bibinfo  {journal} {New J. Phys.}\ }\textbf {\bibinfo {volume}
  {17}},\ \bibinfo {pages} {015006} (\bibinfo {year} {2015})}\BibitemShut
  {NoStop}%
\bibitem [{\citenamefont {Tilloy}\ and\ \citenamefont
  {Di\'osi}(2016)}]{tilloy2016}%
  \BibitemOpen
  \bibfield  {author} {\bibinfo {author} {\bibfnamefont {A.}~\bibnamefont
  {Tilloy}}\ and\ \bibinfo {author} {\bibfnamefont {L.}~\bibnamefont
  {Di\'osi}},\ }\href {\doibase 10.1103/PhysRevD.93.024026} {\bibfield
  {journal} {\bibinfo  {journal} {Phys. Rev. D}\ }\textbf {\bibinfo {volume}
  {93}},\ \bibinfo {pages} {024026} (\bibinfo {year} {2016})}\BibitemShut
  {NoStop}%
\bibitem [{\citenamefont {Altamirano}\ \emph {et~al.}(2016)\citenamefont
  {Altamirano}, \citenamefont {Corona-Ugalde}, \citenamefont {Mann},\ and\
  \citenamefont {Zych}}]{altamirano2016}%
  \BibitemOpen
  \bibfield  {author} {\bibinfo {author} {\bibfnamefont {N.}~\bibnamefont
  {Altamirano}}, \bibinfo {author} {\bibfnamefont {P.}~\bibnamefont
  {Corona-Ugalde}}, \bibinfo {author} {\bibfnamefont {R.~B.}\ \bibnamefont
  {Mann}}, \ and\ \bibinfo {author} {\bibfnamefont {M.}~\bibnamefont {Zych}},\
  }\href {https://arxiv.org/abs/1612.07735} {\bibfield  {journal} {\bibinfo
  {journal} {arXiv:1612.07735}\ } (\bibinfo {year} {2016})}\BibitemShut
  {NoStop}%
\bibitem [{\citenamefont {Tilloy}\ and\ \citenamefont
  {Di\'osi}(2017)}]{tilloy2017}%
  \BibitemOpen
  \bibfield  {author} {\bibinfo {author} {\bibfnamefont {A.}~\bibnamefont
  {Tilloy}}\ and\ \bibinfo {author} {\bibfnamefont {L.}~\bibnamefont
  {Di\'osi}},\ }\href {https://arxiv.org/abs/1706.01856} {\bibfield  {journal}
  {\bibinfo  {journal} {arXiv:1706.01856}\ } (\bibinfo {year}
  {2017})}\BibitemShut {NoStop}%
\bibitem [{\citenamefont {Tilloy}(2017)}]{tilloy2017grw}%
  \BibitemOpen
  \bibfield  {author} {\bibinfo {author} {\bibfnamefont {A.}~\bibnamefont
  {Tilloy}},\ }\href {https://arxiv.org/abs/1709.03809} {\bibfield  {journal}
  {\bibinfo  {journal} {arXiv:1709.03809}\ } (\bibinfo {year}
  {2017})}\BibitemShut {NoStop}%
\bibitem [{\citenamefont {{Bose}}\ \emph {et~al.}(2017)\citenamefont {{Bose}},
  \citenamefont {{Mazumdar}}, \citenamefont {{Morley}}, \citenamefont
  {{Ulbricht}}, \citenamefont {{Toro{\v s}}}, \citenamefont {{Paternostro}},
  \citenamefont {{Geraci}}, \citenamefont {{Barker}}, \citenamefont {{Kim}},\
  and\ \citenamefont {{Milburn}}}]{bose2017}%
  \BibitemOpen
  \bibfield  {author} {\bibinfo {author} {\bibfnamefont {S.}~\bibnamefont
  {{Bose}}}, \bibinfo {author} {\bibfnamefont {A.}~\bibnamefont {{Mazumdar}}},
  \bibinfo {author} {\bibfnamefont {G.~W.}\ \bibnamefont {{Morley}}}, \bibinfo
  {author} {\bibfnamefont {H.}~\bibnamefont {{Ulbricht}}}, \bibinfo {author}
  {\bibfnamefont {M.}~\bibnamefont {{Toro{\v s}}}}, \bibinfo {author}
  {\bibfnamefont {M.}~\bibnamefont {{Paternostro}}}, \bibinfo {author}
  {\bibfnamefont {A.}~\bibnamefont {{Geraci}}}, \bibinfo {author}
  {\bibfnamefont {P.}~\bibnamefont {{Barker}}}, \bibinfo {author}
  {\bibfnamefont {M.~S.}\ \bibnamefont {{Kim}}}, \ and\ \bibinfo {author}
  {\bibfnamefont {G.}~\bibnamefont {{Milburn}}},\ }\href
  {https://arxiv.org/abs/1707.06050} {\bibfield  {journal} {\bibinfo  {journal}
  {arXiv:1707.06050}\ } (\bibinfo {year} {2017})}\BibitemShut {NoStop}%
\bibitem [{\citenamefont {Hall}\ and\ \citenamefont
  {Reginatto}(2017)}]{hall2017}%
  \BibitemOpen
  \bibfield  {author} {\bibinfo {author} {\bibfnamefont {M.~J.~W.}\
  \bibnamefont {Hall}}\ and\ \bibinfo {author} {\bibfnamefont {M.}~\bibnamefont
  {Reginatto}},\ }\href {https://arxiv.org/abs/1706.1707.07974} {\bibfield
  {journal} {\bibinfo  {journal} {arXiv:1707.07974}\ } (\bibinfo {year}
  {2017})}\BibitemShut {NoStop}%
\bibitem [{\citenamefont {Gorini}\ \emph {et~al.}(1976)\citenamefont {Gorini},
  \citenamefont {Kossakowski},\ and\ \citenamefont {Sudarshan}}]{gorini1976}%
  \BibitemOpen
  \bibfield  {author} {\bibinfo {author} {\bibfnamefont {V.}~\bibnamefont
  {Gorini}}, \bibinfo {author} {\bibfnamefont {A.}~\bibnamefont {Kossakowski}},
  \ and\ \bibinfo {author} {\bibfnamefont {E.~C.~G.}\ \bibnamefont
  {Sudarshan}},\ }\href {\doibase 10.1063/1.522979} {\bibfield  {journal}
  {\bibinfo  {journal} {J. Math. Phys.}\ }\textbf {\bibinfo {volume} {17}},\
  \bibinfo {pages} {821} (\bibinfo {year} {1976})}\BibitemShut {NoStop}%
\bibitem [{\citenamefont {Lindblad}(1976)}]{lindblad1976}%
  \BibitemOpen
  \bibfield  {author} {\bibinfo {author} {\bibfnamefont {G.}~\bibnamefont
  {Lindblad}},\ }\href {\doibase 10.1007/BF01608499} {\bibfield  {journal}
  {\bibinfo  {journal} {Commun. Math. Phys.}\ }\textbf {\bibinfo {volume}
  {48}},\ \bibinfo {pages} {119} (\bibinfo {year} {1976})}\BibitemShut
  {NoStop}%
\bibitem [{\citenamefont {Caves}\ and\ \citenamefont
  {Milburn}(1987)}]{caves1987}%
  \BibitemOpen
  \bibfield  {author} {\bibinfo {author} {\bibfnamefont {C.~M.}\ \bibnamefont
  {Caves}}\ and\ \bibinfo {author} {\bibfnamefont {G.}~\bibnamefont
  {Milburn}},\ }\href {\doibase 10.1103/PhysRevA.36.5543} {\bibfield  {journal}
  {\bibinfo  {journal} {Phys. Rev. A}\ }\textbf {\bibinfo {volume} {36}},\
  \bibinfo {pages} {5543} (\bibinfo {year} {1987})}\BibitemShut {NoStop}%
\bibitem [{\citenamefont {Di\'osi}(1990)}]{diosi1990}%
  \BibitemOpen
  \bibfield  {author} {\bibinfo {author} {\bibfnamefont {L.}~\bibnamefont
  {Di\'osi}},\ }\href {\doibase 10.1103/PhysRevA.42.5086} {\bibfield  {journal}
  {\bibinfo  {journal} {Phys. Rev. A}\ }\textbf {\bibinfo {volume} {42}},\
  \bibinfo {pages} {5086} (\bibinfo {year} {1990})}\BibitemShut {NoStop}%
\bibitem [{\citenamefont {Wiseman}\ and\ \citenamefont
  {Milburn}(1993)}]{wiseman1993}%
  \BibitemOpen
  \bibfield  {author} {\bibinfo {author} {\bibfnamefont {H.~M.}\ \bibnamefont
  {Wiseman}}\ and\ \bibinfo {author} {\bibfnamefont {G.~J.}\ \bibnamefont
  {Milburn}},\ }\href {\doibase 10.1103/PhysRevLett.70.548} {\bibfield
  {journal} {\bibinfo  {journal} {Phys. Rev. Lett.}\ }\textbf {\bibinfo
  {volume} {70}},\ \bibinfo {pages} {548} (\bibinfo {year} {1993})}\BibitemShut
  {NoStop}%
\bibitem [{\citenamefont {Wiseman}(1994)}]{wiseman1994}%
  \BibitemOpen
  \bibfield  {author} {\bibinfo {author} {\bibfnamefont {H.~M.}\ \bibnamefont
  {Wiseman}},\ }\href {\doibase 10.1103/PhysRevA.49.2133} {\bibfield  {journal}
  {\bibinfo  {journal} {Phys. Rev. A}\ }\textbf {\bibinfo {volume} {49}},\
  \bibinfo {pages} {2133} (\bibinfo {year} {1994})}\BibitemShut {NoStop}%
\bibitem [{\citenamefont {Diosi}\ and\ \citenamefont
  {Gisin}(1994)}]{diosi1994}%
  \BibitemOpen
  \bibfield  {author} {\bibinfo {author} {\bibfnamefont {L.}~\bibnamefont
  {Diosi}}\ and\ \bibinfo {author} {\bibfnamefont {N.}~\bibnamefont {Gisin}},\
  }\href {\doibase 10.1103/PhysRevLett.72.4053} {\bibfield  {journal} {\bibinfo
   {journal} {Phys. Rev. Lett.}\ }\textbf {\bibinfo {volume} {72}},\ \bibinfo
  {pages} {4053} (\bibinfo {year} {1994})}\BibitemShut {NoStop}%
\bibitem [{\citenamefont {Wiseman}\ and\ \citenamefont
  {Milburn}(2009)}]{wiseman2009}%
  \BibitemOpen
  \bibfield  {author} {\bibinfo {author} {\bibfnamefont {H.~M.}\ \bibnamefont
  {Wiseman}}\ and\ \bibinfo {author} {\bibfnamefont {G.~J.}\ \bibnamefont
  {Milburn}},\ }\href@noop {} {\emph {\bibinfo {title} {Quantum measurement and
  control}}}\ (\bibinfo  {publisher} {Cambridge university press, Cambridge
  UK},\ \bibinfo {year} {2009})\BibitemShut {NoStop}%
\bibitem [{\citenamefont {Jacobs}\ and\ \citenamefont
  {Steck}(2006)}]{jacobs2006}%
  \BibitemOpen
  \bibfield  {author} {\bibinfo {author} {\bibfnamefont {K.}~\bibnamefont
  {Jacobs}}\ and\ \bibinfo {author} {\bibfnamefont {D.~A.}\ \bibnamefont
  {Steck}},\ }\href {\doibase 10.1080/00107510601101934} {\bibfield  {journal}
  {\bibinfo  {journal} {Contemporary Physics}\ }\textbf {\bibinfo {volume}
  {47}},\ \bibinfo {pages} {279} (\bibinfo {year} {2006})}\BibitemShut
  {NoStop}%
\bibitem [{\citenamefont {Di{\'o}si}(2003)}]{diosi2003}%
  \BibitemOpen
  \bibfield  {author} {\bibinfo {author} {\bibfnamefont {L.}~\bibnamefont
  {Di{\'o}si}},\ }\enquote {\bibinfo {title} {Progressive decoherence and total
  environmental disentanglement},}\ in\ \href {\doibase
  10.1007/3-540-44874-8_8} {\emph {\bibinfo {booktitle} {Irreversible Quantum
  Dynamics}}},\ \bibinfo {editor} {edited by\ \bibinfo {editor} {\bibfnamefont
  {F.}~\bibnamefont {Benatti}}\ and\ \bibinfo {editor} {\bibfnamefont
  {R.}~\bibnamefont {Floreanini}}}\ (\bibinfo  {publisher} {Springer Berlin
  Heidelberg},\ \bibinfo {address} {Berlin, Heidelberg},\ \bibinfo {year}
  {2003})\ pp.\ \bibinfo {pages} {157--163}\BibitemShut {NoStop}%
\bibitem [{\citenamefont {Yu}\ and\ \citenamefont {Eberly}(2004)}]{yu2004}%
  \BibitemOpen
  \bibfield  {author} {\bibinfo {author} {\bibfnamefont {T.}~\bibnamefont
  {Yu}}\ and\ \bibinfo {author} {\bibfnamefont {J.~H.}\ \bibnamefont
  {Eberly}},\ }\href {\doibase 10.1103/PhysRevLett.93.140404} {\bibfield
  {journal} {\bibinfo  {journal} {Phys. Rev. Lett.}\ }\textbf {\bibinfo
  {volume} {93}},\ \bibinfo {pages} {140404} (\bibinfo {year}
  {2004})}\BibitemShut {NoStop}%
\bibitem [{\citenamefont {Wong}\ and\ \citenamefont {Zakai}(1965)}]{wong1965}%
  \BibitemOpen
  \bibfield  {author} {\bibinfo {author} {\bibfnamefont {E.}~\bibnamefont
  {Wong}}\ and\ \bibinfo {author} {\bibfnamefont {M.}~\bibnamefont {Zakai}},\
  }\href {\doibase 10.1214/aoms/1177699916} {\bibfield  {journal} {\bibinfo
  {journal} {Ann. Math. Statist.}\ }\textbf {\bibinfo {volume} {36}},\ \bibinfo
  {pages} {1560} (\bibinfo {year} {1965})}\BibitemShut {NoStop}%
\end{thebibliography}%
\bibliographystyle{apsrev4-1}
\end{document}